\documentclass[prb,preprint,eqsecnum]{revtex4}
\usepackage{amsmath}
\usepackage{graphicx}
\usepackage{amssymb}
\usepackage{graphicx}
\usepackage{amsfonts}

\begin{document}

\title{Flux melting in BSCCO: Incorporating  both
  electromagnetic and Josephson couplings}

\author{Sandeep Tyagi}
\author{Yadin Y. Goldschmidt}
\affiliation{Department of Physics and Astronomy, University of Pittsburgh,
Pittsburgh, Pennsylvania 15260}


\begin{abstract}
  
  Multilevel Monte Carlo simulations of a BSCCO system are carried out
  including both Josephson as well as electromagnetic couplings for a
  range of anisotropies. A first order melting transition of the flux
  lattice is seen on increasing the temperature and/or the magnetic
  field.  The phase diagram for BSCCO is obtained for different values
  of the anisotropy parameter $\gamma$. The best fit to the experimental
  results of D. Majer {\it et al.} [Phys. Rev. Lett. {\bf 75}, 1166
  (1995)] is obtained for $\gamma\approx 250$ provided one assumes 
  a temperature dependence $\lambda^2(0)/\lambda^2(T)=1-t$ of the
  penetration depth with $t=T/T_c$. Assuming a dependence 
  $\lambda^2(0)/\lambda^2(T)=1-t^2$ the best fit is obtained for $
  \gamma\approx 450$. 
  For finite anisotropy the data is shown to collapse on a straight
  line when plotted in dimensionless units which shows that the
  melting transition can be satisfied with a single Lindemann
  parameter whose value is about 0.3. A different scaling applies to
  the $\gamma=\infty$ case.
  The energy jump is measured across the transition and for large values of
  $\gamma$ it is found to increase with
  increasing anisotropy and to decrease with increasing magnetic field.
  For infinite anisotropy we see a 2D behavior of flux droplets with a
  transition taking place at a temperature independent of the magnetic
  field.  We also show that for smaller values of anisotropy 
  it is reasonable to replace the electromagnetic
  coupling with an in-plane interaction represented by a Bessel
  function of the second kind ($K_0$), thus justifying our claim in a
  previous paper.

\end{abstract}

\maketitle

\section{Introduction}

High-temperature superconductors are materials of type II, that allow
for partial magnetic flux penetration if the external field satisfies
$H_{c1}<H<H_{c2}$ \cite{tinkham,blatter,brandt}. The flux penetrates
the sample in the form of flux-lines (FL's), each containing a quantum unit
$\phi_0=hc/2e$ of flux. At low temperature the FL's form an ordered
hexagonal lattice (Abrikosov lattice) due to their their mutual
repulsion. The lattice constant is given by
$a_0=\sqrt{2\phi_0/\sqrt{3}B}$, where $B$ is the magnetic field flux density.
At high temperature and/or magnetic field this lattice
melts due to thermal fluctuations \cite{exp1,exp2,cubitt,exp3,exp4}.
Two of the most commonly used high-temperature superconductors are
YBa$_2$Cu$_3$O$_{7-\delta}$
(YBCO) and  Bi$_2$Sr$_2$CaCu$_2$O$_{8+\delta}$ (BSCCO). They
are very different in one very important respect - their anisotropy
parameter $\gamma$, defined as $\gamma^2=m_z/m_\perp$,
where $m_z$ and $m_\perp$ denote 
the effective masses of electrons moving along the $c$ axis and the
$ab$ plane, respectively. While
for YBCO the anisotropy is somewhere between 5-7, for BSCCO
it is estimated to be between 10 to a 100 times larger, and estimates
vary across the literature. What contributes to the discrepancy of
different estimates is the fact that the anisotropy varies depending on the
degree of doping. There is also the question of how to extract the
value of anisotropy correctly from the exerimental results.
Blatter {\it et al} \cite{blatter} cite a range of anisotropies of
50-200 for BSCCO. References [\onlinecite{exp3,martinez,lee}] cite values in
the range of 140-160. However more recent acurate measurements 
\cite{gaifullin}
the components of the London penetration depths $\lambda_c$ and
$\lambda_{ab}$, the ratio of which is $\gamma$, are consistent with
anisotropy in the range of 300-500 for optimally doped samples. The
previously reported lower values may belong to overdoped samples or
constituted only a lower bound.
Anisotropy controls the amount of ``wiggling'' of a flux-line from 
plane to plane. In YBCO the FL's are more rigid while in BSCCO they
are so loose that they are customarily referred to as a stack of two
dimensional pancakes \cite{clem} (or droplet vortices) rather than FL's. 

Interaction between two FL's in YBCO is 
non-local. It is a screened Bio-Savart type of interaction where each segment
of a FL  interacts with every other segment of the same FL and all the 
other FL's. For segments oriented in the same
direction the interaction is repulsive \cite{blatter}: 
\begin{eqnarray}
  {\cal F}=\frac{\varepsilon_0}{2}\sum_{ij}\int d{\mathbf
    s}_i(z) \cdot 
  d{\mathbf s}_j(z') \frac{\exp(-|{\mathbf s}_i(z)-{\mathbf
        s}_j(z')|/\lambda)}{|{\mathbf s}_i(z)-{\mathbf s}_j(z')|} 
\end{eqnarray}
Here $\mathbf{s}_i(z)$  denotes the position of the $i$'th FL at
elevation $z$ along the $z$-axis, $\varepsilon_0=\phi_0^2/(4\pi\lambda)^2$
is the line energy and $\lambda$ is the screening length (penetration
depth). 
Considering two given FL's, it 
turns out though that it is a good approximation to replace the non-local 
interaction of a given line segment of one FL, with all
segments of the other FL with a single
interaction among segments belonging to the same plane. This 
interaction is given approximately by \cite{nordborg} $2 \,\varepsilon_0 \,d\,
K_0(R_{ij}/\lambda)$ where  
$R_{ij}$ is the distance between the two segments in the same plane and
$d$ is the thickness of each layer.
The approximation is valid when the FL's do not deviate too much
from straight lines which is a good approximation for YBCO in the
``solid'' vortex lattice phase, because FL's are stiff and do not
wiggle too much. 
For each FL, there is also an elastic energy
associated with its deviation from a straight line along the
$z$-direction. The elastic energy of a flux-line in YBCO is
approximately given by   
\begin{eqnarray} 
\frac{\epsilon_l}{2}\int_0^L dz \,(d{\mathbf R}(z)/dz)^2,  
\end{eqnarray}
assuming the external magnetic field is aligned along the z-direction. 
The elastic coefficient (line tension) $\epsilon_l$ is equal to
$\varepsilon_0 \,\ln (\lambda/\xi)/\gamma^2$ where $\xi$ is the coherence length, and $\gamma$ is the anisotropy. 
In the discrete case this self-energy
transforms into an attractive quadratic interaction between segments
in adjacent planes. In this form the problem is equivalent to a system
of bosons with repulsive interactions \cite{nelson,nordborg}. The term
described in the last equation corresponds to the kinetic energy of
the bosons which repel each other with a screened Coulomb interaction.

For BSCCO the situation is different because each FL is
represented more faithfully by a collection of pancakes. Each pancake
interacts with every other pancake, but the interaction is different
from the interaction among FL segments discussed above. The
interaction can be shown to consist of two parts. The first part is
called the electromagnetic interaction (or simply magnetic) and it
exists even in the case that 
the layers of the materials are completely decoupled, so no current
can flow along the $c$-axis of the sample. A pancake vortex located in one
plane gives rise to screening currents in the same plane as well
as in all other planes. A second pancake vortex, located elsewhere,
interacts  with the screening currents induced by the first
pancake \cite{artemenko}. This interaction has been calculated by Clem 
and others \cite{clem}. Two pancakes in the same plane interact with a
repulsive 
interaction while pancakes in different planes attract one another.
If one considers a single pancake vortex and an infinite set of
pancakes a distance $R$ away stacked along the $z$-axis, then the
interaction still sums to $2 \,\varepsilon_0 \,d \,K_0(R/\lambda)$ in the
limit when $d/\lambda$ goes to zero (see Appendix A).
Clem \cite{clem} proceeds to show that if one has a
straight array of pancake vortices along the $z$-axis, and one pancake
of the stack is displaced a distance $R$ in the lateral direction then
the magnetic energy of the configuration increases by an amount
\begin{equation}
\Delta E(R)=  \frac{d{\phi _{0}}^{2}}{8{\pi }^{2}{\lambda }^{2}}\left(
  {\cal C}+\ln\left(\frac{R}{2
      \lambda}\right)+K_0\left(\frac{R}{\lambda}\right)\right). 
\end{equation}
where ${\cal C}$ is Euler's constant (=0.5772...). For large $R$
($R\gg\lambda$), the 
modified Bessel function $K_0$ decays exponentially and thus the
energy increases like $\ln(R/\lambda)$. For small $R$ the Bessel
function can be expanded in a power series in $R/\lambda$ 
\begin{equation}
  K_0(R/\lambda)=-\ln(R/2\lambda)(1+R^2/4 \lambda^2+\cdots)-{\cal
    C}+R^2 (1-{\cal C})/4 \lambda^2+\cdots,
\end{equation}
and thus the electromagnetic energy behaves like $R^2$ to leading
order in $R$.

The second part of the interaction among pancake vortices is the
so-called Josephson interaction \cite{artemenko,clem1,blatter}.
It results from the fact that there is a
Josephson current flowing between two superconductors separated by an
insulator and this current is proportional to the sine of the phase
difference of 
the superconducting wave functions. The superconductors in the present
case are the different CuO$_2$ planes. When two pancakes belonging to
the same stack and residing in adjacent
planes move away from each other, the phase difference that originates
causes a Josephson current to begin flowing between the planes. This
results in an attractive interaction between pancakes that for
distances small compared to $r_g\equiv \gamma d$ is approximately
quadratic \cite{artemenko,blatter} in the distance. When the two adjacent
pancakes are separated by a distance larger than $r_g$, a ``Josephson
string'' is formed, whose energy is proportional to its length \cite{clem1}. 

When the anisotropy is not too large, the
Josephson coupling among adjacent pancakes, which are loosely belonging to the
same ``flux-line'', dominates over the electromagnetic interaction, and
the later can be neglected. The ratio
of the coefficients of the quadratic terms in the effective
electromagnetic interaction (as mentioned above) and the
Josephson interaction goes roughly like $\gamma^2(d/\lambda)^2$ (where
$d/\lambda \sim 1/120$ for BSCCO at $T=0$ and even smaller at higher
temperatures). Thus for anisotropy 
$\gamma=50$ we get a factor of 0.25 or less (a somewhat more precise estimate
\cite{ryu} gives a ratio of about 0.1). Thus the magnetic interaction is
small compared to the Josephson interaction for anisotropies in
the range of $\gamma=50-100$. For
samples with $\gamma=200$ these interactions are already comparable. For large
values of $R$ the magnetic interaction increases logarithmically and
the Josephson interaction increases linearly so the electromagnetic
interaction is always negligible. The key to the estimate given above is
to consider not just two pancake vortices but a whole line with one
displaced pancake. This argument is valid if the deviations of the
vortices from straight lines are not too large.
On the other hand, the electromagnetic
interaction starts to be important for anisotropies which are significantly
larger than $\lambda/d$ which for BSCCO is about 120.

It is the aim of this paper to include both the electromagnetic
interaction and the Josephson interaction among pancakes and to see
what is the combined effect on the phase diagram of the melting
transition and the energy jump across the transition.
The electromagnetic interaction will be included fully in the sense
that we will not make the approximation that the pancake stacks are
nearly straight and hence the electromagnetic coupling will not be
replaced by an in-plane effective coupling.
Numerically, much of the past work on BSCCO has been confined to the
X-Y model \cite{hu,hu1,koshelev,nonomura,olsson,chin}
and Bose model \cite{ryu,sandeep}, both of them treat the electromagnetic
coupling imprecisely
by including it as an effective in-plane interaction.
Recently in several papers using the Langevin simulation method 
\cite{olson,kolton,reichhardt},
the electromagnetic coupling has been fully taken into account. 
Unfortunately, these papers completely neglect the
Josephson coupling which can hardly be justified. Also some of these
papers work with a small system size like $5$ to $10$ planes along the
z-direction, and some do not even
use periodic boundary conditions in the z-direction. In this paper we
carry out Monte Carlo simulations of a BSCCO system consisting of $20-36$
planes and periodic boundary conditions are used in all directions,
including the 
$z$ direction. Thus a pancake would interact with an infinite number
of pancakes through the images under the periodic boundary
conditions. For this an efficient way to sum over the interaction is
required. We derive a formula for summing over the logarithmic interaction in
Appendix B.

\section{The model}
The starting point is the Lawrence-Doniach \cite{ld} Gibbs free-energy
functional, 
\begin{eqnarray}
\mathcal{G}[\psi _{n},\mathbf{a}] &=&\int d^{2}\mathbf{R}
\,\sum_{n}\alpha |\psi
_{n}|^{2}+\frac{\beta }{2}|\psi _{n}|^{4}+\frac{\hbar ^{2}}{2m}\left\vert
\left( \frac{\mathbf{\nabla }^{(2)}}{i}+\frac{2\pi }{\phi _{0}}\mathbf{a}%
^{(2)}\right) \psi _{n}\right\vert ^{2}  \notag \\
&&+\frac{\hbar ^{2}}{2Md^{2}}\left\vert \psi _{n+1}\exp \left( \frac{2\pi i}{%
\phi _{0}}\int_{n\,d}^{(n+1)d}dz\,a_{z}\right) -\psi _{n}\right\vert ^{2} 
\notag \\
&&+\int d^{2}\mathbf{R} dz\left( \frac{b^{2}}{8\pi }-\frac{\mathbf{b}\cdot 
\mathbf{H}}{4\pi }\right) ,
\end{eqnarray}
where $\psi _{n}~$represents the superconducting order parameter in the $%
n^{th}$ CuO$_2$ layer,
$\mathbf{a}^{(2)}$ is the vector potential in the plane, and  
$d$ is the thickness of the insulating layers. $\mathbf{b}$ is the
local magnetic field and $\mathbf{H}$ the externally applied field. The
usual 3D integration of the GL theory has been replaced by a summation over all
the superconducting layers along with a 2D integration over the superconducting
planes.
We set $\psi _{n}=|\psi _{n}|\exp (i\phi _{n})$ and, working in the London
approximation, we drop the term $\ \alpha |\psi _{n}|^{2}+\beta
|\psi _{n}|^{4}/2$ because it gives only a constant     
contribution. Then we get  
\begin{eqnarray}
\mathcal{G}=\int d^{2}\mathbf{R} \,\frac{\varepsilon _{0}d}{2\pi }\left( 
\int dz~\sum_{n}\delta _{2}(z-n\,d)\left( \frac{\mathbf{\nabla }^{(2)}\phi
_{n}}{i}+\frac{2\pi }{\phi _{0}}\mathbf{a}^{(2)}\right) ^{2} \notag
\right. &&\\
\left. +\frac{2m}{Md^{2}}\sum_{n}
\left[ 1-\cos \left( \phi _{(n+1)}-\phi _{n}+\frac{%
2\pi }{\phi _{0}}\int_{nd}^{(n+1)d}dz\,a_{z}\right) \right]%
\right) \notag \\ 
+\int d^{2}\mathbf{R}dz\left( \frac{b^{2}}{8\pi }-\frac{\mathbf{b\cdot
      H}}{4\pi}\right), 
\label{ldmodel}
\end{eqnarray}
where
\begin{equation}
\varepsilon _{0}= 2\pi \frac{\hbar ^{2}|\psi _{n}|^{2}}{2m}=\frac{\phi
  _{0}^{2}}{(4\pi \lambda )^{2}}. 
\end{equation}
 Minimization with respect to $a_{x}$ (i.e. $\frac{\delta \mathcal{G}}{\delta a_{x}}=0$) and $%
a_{y}$ (i.e. $\frac{\delta \mathcal{G}}{\delta a_{y}}=0$) gives
\begin{equation}
\lambda ^{2}\mathbf{\triangle a}^{(2)}=d\sum_{n}\delta _{2}(z-n\,d)\left[ 
\mathbf{a}^{(2)}+\frac{\phi _{0}}{2\pi }\mathbf{\nabla }^{(2)}\phi _{n}%
\right] ,  \label{5afield}
\end{equation}
where $\mathbf{\triangle }$ stands for the 3-dimensional Laplacian.
Minimization with respect to $a_{z}$ gives
\begin{equation}
\frac{\varepsilon _{0}d}{2\pi }\frac{2m}{Md^{2}}\frac{
2\pi }{\phi _{0}}\sin (\Phi _{(n,n+1)})+\frac{1}{4\pi }\left(
\mathbf{\nabla }\times \left(  
\mathbf{\nabla }\times \mathbf{a}\right) \right) _{z}=0,
\label{2deq}
\end{equation}
where 
\begin{equation}
\Phi_{n+1,n}=\phi _{n+1}-\phi _{n}+\frac{2\pi }{\phi _{0}}%
\int_{nd}^{(n+1)d}dz~a_{z}
\end{equation}
is the gauge invariant phase difference between the layers $n$ and
$n+1$.
Eq. (\ref{2deq}) implies 
\begin{equation}
\Delta a_{z}=\frac{4\pi }{c}j_{J}\sin (\Phi_{n,n+1}).
\label{josephson1}
\end{equation}
where 
\begin{eqnarray}
  j_J=\frac{c\phi_0}{8\pi^2\lambda^2\gamma^2 d}
\end{eqnarray}
is the Josephson-coupling current density between layers. Minimization
with respect to $\phi _{n}$ gives 
\begin{equation}
 \Delta^{(2)}\phi _{n}+\frac{2\pi }{\phi _{0}}%
\mathbf{\nabla }^{(2)}\cdot \mathbf{a}^{^{(2)}} =\frac{1}{\gamma
  ^{2} d^2}\left[\sin (\Phi_{n,n-1})-\sin (\Phi_{n+1,n})\right].
\label{ff1}
\end{equation}
Eqs. (\ref{5afield}), (\ref{josephson1}) and (\ref{ff1}) are to be
solved with the appropriate boundary conditions and the solution must
be substituted back into Eq. (\ref{ldmodel}) to obtain the Gibbs
free-energy, and thus the strength of the interaction among pancake
solutions. We also see that in the limit of infinite anisotropy the
right-hand-side of equations (\ref{josephson1}) and (\ref{ff1}) tend to
zero. An isolated pancake residing in plane $n$ is a singular solution of the
equation for the phase of the wave function which satisfies
\begin{equation}
\mathbf{\nabla }^{(2)}\phi _{n}(\mathbf{R })=-\frac{\mathbf{n}\times
\left({\mathbf{R }}-{\mathbf{R}}_{n}\right)}{\left({\mathbf{R
    }}-{\mathbf{R}}_{n}\right)^2 }, 
\end{equation}
where $\mathbf{R}$ is a two dimensional vector in the plane and
$\mathbf{R}_n$ denotes the center of the pancake.  By
$\mathbf{n}$ we denote a unit vector in the z-direction. Thus as one
fully encircles the pancake the phase $\phi_n$ 
changes by $2\pi$, and is singular at the center of the pancake. In
the case when the Josephson coupling is totally neglected, i.e. for
$\gamma\rightarrow\infty$, the
full solution of Eqs. (\ref{5afield}), (\ref{josephson1}) and (\ref{ff1})
can be found and  from it one can easily obtain the magnetic-field in
real space \cite{clem}. 
It is not a trivial matter to switch from the variables $\phi_n$ and
$\mathbf{a}$ to pancake variables and express the free energy in terms
of the latter. This transformation can only be implemented
approximately. In the following we first summarize the known results for the
Josephson interaction and the electromagnetic interaction and then
proceed to combine them together into a single algorithm. Most papers
consider only one type of interaction, Josephson or electromagnetic in
the limit that one dominates over the other.

\subsection{Josephson Coupling}
We keep the same Josephson coupling as in
Ref. \onlinecite{sandeep}. This coupling 
is strongly dependent on the anisotropy parameter $\gamma$.
Consider two adjacent pancakes, belonging to the same FL, residing in the $m$
and $m+1$ planes respectively, such that their centers are displaced
and not located on top of each other. Assume that the pancake on the
$m+2$'th plane is located on top of the $m+1$-plane pancake, and the
pancake residing in the $m-1$ plane is at the same position as the
$m$-plane pancake. Thus $\phi_{m+1}\neq \phi_m$ but $\phi_{m+2}=
\phi_{m+1}$ and $\phi_{m-1}= \phi_m$. This 
assumption is made in order to trancate the infinite set of couples equations
\cite{blatter}, and in real situations may constitue an approximation. 
Denoting $\Phi_{m+1,m}=\phi_{m+1}-\phi_m$ simply 
by $\Phi$ we see that writing down Eq.~(\ref{ff1}) for the $m$ and
$m+1$ planes respectively and subtracting one equation from the other,
one obtains
\begin{eqnarray}
\mathbf{\Delta }^{(2)}\Phi=\frac{2}{r_g^2}\sin(\Phi),  
\label{SG}
\end{eqnarray}
where $r_{g}=\gamma d$ is the relevant screening length of the problem. Note
that the screening term due to the vector potential has been neglected
since it is negligible on length scales $R\sim r_g \ll \gamma \lambda$. 
Equation (\ref{SG}) is the famous Sine Gordon Equation. Once its
solution is obtained, it needs to be substituted in the
Lawrence-Doniach Gibbs free-energy. This results in a contribution of
the form \cite{feigelman}
\begin{eqnarray}
  \mathcal{G}_J = \frac{d \varepsilon_0}{\pi r_g^2} \int
  d^2\mathbf{R}'[1-\cos(\Phi(\mathbf{R}')).
\label{jfe}
\end{eqnarray}
If we denote the separation between the two pancakes by $R$
one can show that for $R\ll R' \ll r_g$ the solution of Eq.~(\ref{SG})
is given simply by 
\begin{eqnarray}
  \Phi(\mathbf{R}')=R \sin(\theta')/R',
\label{phi1}
\end{eqnarray}
since the right-hand-side can be neglected in this region. Here
$\theta'$ is the azimuthal angle in the plane. For $R' \gg r_g$ the
$\sin(\Phi)$ can be replaced by $\Phi$ and we see that the
solution decays exponentially with a screening length $r_g$. 
Thus substituting Eq.~(\ref{phi1}) into
Eq.~(\ref{jfe}), expanding the cosine to quadratic order and cutting
off the integration at a large distance $R'=r_g$ and small distance $R'=R$
the interaction energy becomes
\begin{eqnarray}
\mathcal{G}_J (R)=\frac{d \varepsilon_0}{2} 
\ln\left(\frac{r_g}{R}\right)\left(\frac{R}{r_g}\right)^2, 
\label{gj1}
\end{eqnarray}
so we see that it is approximately proportional to $R^2$. On the other
hand when the 
separation between the centers of the two pancakes becomes larger than
$r_g$ then a Josephson string is formed between the $m$'th and the
$m+1$'th planes in a direction parallel to the planes. The energy of
the Josephson string is proportional to its length which is equal to
$R$, the pancakes' separation. The calculation of this energy is
rather involved and discussed by Clem, Coffey and Hao \cite{clem1}. The
final result is
\begin{eqnarray}
  \mathcal{G}_J(R)={d\varepsilon_0}\left(1.12+\ln\left(\frac{\lambda}{d}
\right)\right)\frac{R}{r_g}.
\label{gj2}
\end{eqnarray}
The question now arises how to match these two interaction potentials
valid for $R \ll r_g$ and $R \gg r_g$, one behaving quadratically in
$R$ and one linearly in $R$. One such extrapolation was
given by Ryu, Doniach, Deutscher and Kapitulnik (RDDK)\cite{ryu}, who
achieved a matching by keeping the coefficient of the linear term in
$R$ as given in Eq.~(\ref{gj2}) in both regions, choosing the matching
point to be at $R=2r_g$, and subtracting a constant so that the two
expressions would vanish at the matching point. Keeping the same
constant in both expressions assures that the first derivative is
continuous at the matching point.
RDDK also replaced the constant 1.12 by 1. There are ways to
improve the extrapolation \cite{yygst} but they do not change
significantly the results of the simulations performed using the RDDK
formula. Unfortunately there was a mistake by a factor of $\pi/2$ in
the RDDK formula that we corrected below that has the effect of
renormalizing the anisotropy parameter by about $\sqrt{\pi/2}\approx 
1.25$ since 
the major contribution to the simulations come from the region $R<2r_g$. 
To summarize,
the London free-energy for inter-layer (IL) Josephson coupling is
given approximately by: 
\begin{equation}
\Im _{IL}(\mathbf{R}_{i,m},\mathbf{R}_{i,m+1})=\frac{d{\phi
    _{0}}^{2}}{16{\pi }^{2} 
{\lambda }^{2}}\left(1+\ln \left(\frac{\lambda }{d}\right)\right)
\left[\frac{{(|\mathbf{R}_{i,m}-\mathbf{R}_{i,m+1}|})^{2}}
{4{r_{g}}^{2}}-1\right],
\label{quadjos}
\end{equation}
 for $|\mathbf{R}_{i,m}-\mathbf{R}_{i,m+1}|<2r_{g}$ , and 
\begin{equation}
\Im _{IL}(\mathbf{R}_{i,m},\mathbf{R}_{i,m+1})=\frac{d{\phi
    _{0}}^{2}}{16{\pi }^{2} 
{\lambda }^{2}}\left(1+\ln \left(\frac{\lambda }{d}\right)\right)
\left[\frac{|\mathbf{R}_{i,m}-\mathbf{R}_{i,m+1}|}{r_{g}}-2\right],
\label{linjos}
\end{equation}
for $|\mathbf{R}_{i,m}-\mathbf{R}_{i,m+1}|>2r_{g}$. Here the position
of a pancake is specified in 
terms of cylindrical coordinates. Thus the position of the $i$'th
pancake in the $m$'th plane is given by $(\mathbf{R}_{i,m},\ md)$,
where $\mathbf{R}_{i,m}$ is a two 
dimensional vector in the {\it ab} plane. 
The index $i$ labels the FL that the pancake is a part of. We
have considered only pancakes belonging to the same FL. This is
because for large separations by definition the Josephson string is
formed among pancakes belonging to the same FL. We do check for the
energy of all nearest neighboring pairs when deciding how to connect
pancakes and 
we allow the process of flux ``cutting'' and switching. In the case that
pancakes belonging to different FL's approach each other to a distance
much smaller than $r_g$, than there will be an interaction of the form of
Eq.~(\ref{gj1}), however this occurrence is rather rare for magnetic
fields of the strength considered here and hence neglected by RDDK and
also in this work.

When using Eqs.(\ref{quadjos}) and (\ref{linjos}) it is necessary to
specify the temperature dependence of $\lambda$. Different choices for
this dependence are found in the literature. In this work we used
the same choice as in Refs. \cite{ryu,blatter,magro}, motivated by 
Ginzburg-Landau (GL):
\begin{eqnarray}
  \frac{\lambda^2(0)}{\lambda^2(T)}={1-T/T_c}.
\label{lambdaT1}
\end{eqnarray}
Some authors \cite{anisotropy,dodgson} use a dependence of the form
\begin{eqnarray}
  \frac{\lambda^2(0)}{\lambda^2(T)}=1-(T/T_c)^2.
\label{lambdaT2} 
\end{eqnarray}

\begin{figure}
\includegraphics[width=4in,height=3in]{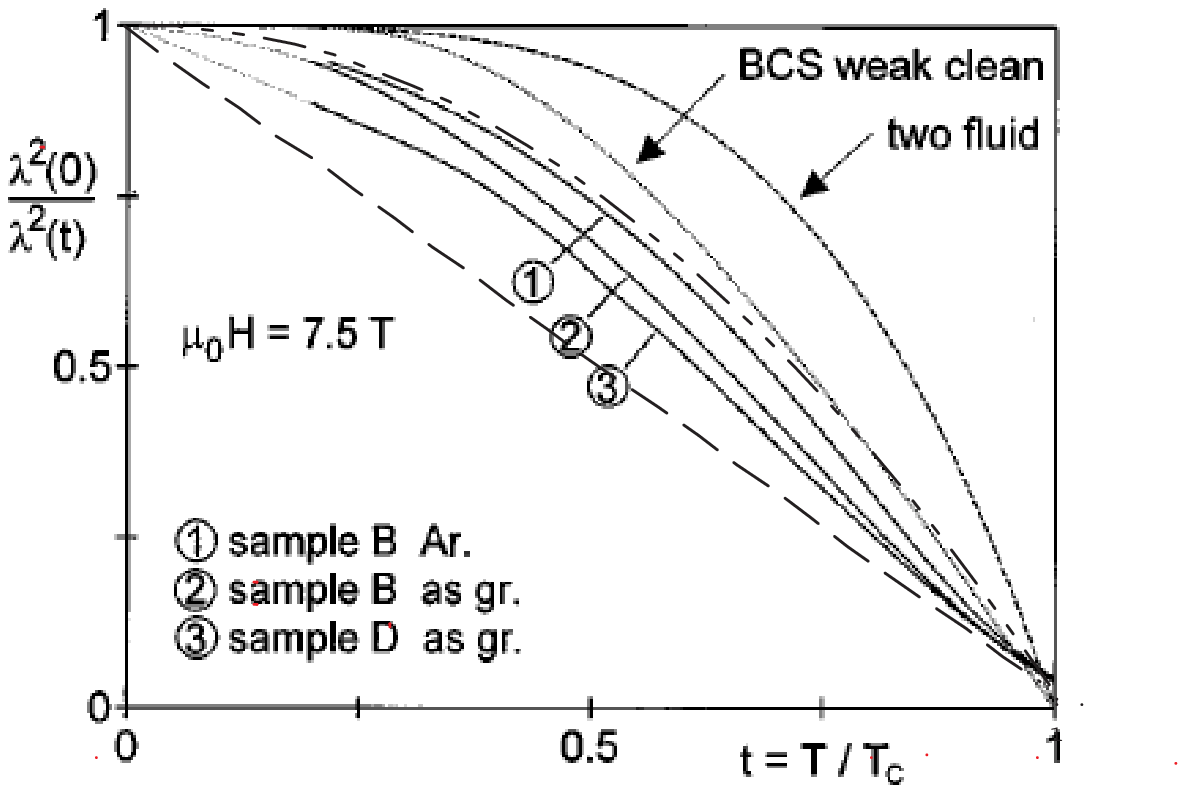}
\caption{Temperature dependence of the normalized penetration depth
for 3 experimental samples as given in Ref. \onlinecite{waldmann}. For
comparison the two-fluid and BCS result for clean superconductors in
the weak coupling limit are shown. We also added the linear and
quadratic behavior as given in Eqs. (\ref{lambdaT1}) and
(\ref{lambdaT2}) in the text, represented by dashed and dot-dashed
lines respectively. }  
\label{lambda}
\end{figure}

Recent experiments \cite{waldmann} show that 
the temperature dependence of the london penetration depth is not
universal and depend on the amount
of doping and the sample history. In Fig. (\ref{lambda}) we show 
the temperature dependence of the three samples investigated in
Ref.[\onlinecite{waldmann}]. Also shown on the figure is the two fluid
dependence \cite{tinkham} ${\lambda^2(0)}/{\lambda^2(T)}=1-(T/T_c)^4$
which normaly applies in the opposite limit of very small GL ratio
$\kappa=\lambda(0)/\xi(0)$, unlike high-Tc materials. Displayed also
is the weak  
coupling, clean limit BCS curve. Both curves are lying above the
experimental data. We added 
to the original figure the two behaviors given in
Eqs. (\ref{lambdaT1}) and (\ref{lambdaT2}) to show that the
experimental data actually falls between theses two curves, so they
give reasonable lower and upper bounds to the experimental
results. In the result section we
discuss how this choice of temperature dependence affects the 
comparison of the simulation results with experiments.

\subsection{Electromagnetic Coupling}
This coupling can be obtained from the Lawrence-Doniach model discussed at the
beginning of this section by putting $\gamma=\infty$, which eliminates
the Josephson coupling altogether. Extensive calculations can be found
in the literature \cite{clem,blatter,brandt}.
For the in-plane interaction between two pancakes one
finds,  
\begin{eqnarray}
{\bf U}(R_{ij},0)=2d\varepsilon_{0}
\left(\left(1-\frac{d}{2\lambda}\right)\ln{\frac{C}{R_{ij}}}
+\frac{d}{2\lambda}
E_{1}\right) \ ,
\label{electromagnetic1}
\end{eqnarray}
where $R_{ij}=| \mathbf{R}_{i,m}-\mathbf{R}_{j,m}|$  is the radial
distance in cylindrical coordinates and  ``$m$'' denotes
the index of the plane.  

The interaction between two pancakes
situated at different planes $(\mathbf{R}_{i,m},\ md)$ and
$(\mathbf{R}_{j,n},\ nd)$  
is given by,
\begin{eqnarray}
{\bf U}(R_{ij},z)=-\frac{d^{2}\varepsilon_{0}}{\lambda}
\left(\exp(-|z|/\lambda)\ln\frac{C}{R_{ij}}-
E_{2}\right) \ , 
\label{electromagnetic2}
\end{eqnarray}
where $R_{ij}= |\mathbf{R}_{i,m}-\mathbf{R}_{j,n}|$, and $z=(m-n) d$.

In the above equations we defined 
\begin{eqnarray}
  E_{1} = 
\int^{\infty}_{R_{ij}} d\rho
\exp(-\rho/\lambda)/\rho,\\ 
E_{2} = 
\int^{\infty}_{R_{ij}} d\rho \exp(-\sqrt{z^{2}+\rho
   ^2} /\lambda)/\rho,
\label{e1e2}
\end{eqnarray} 
$C$ is a constant of the order of the system's size
that cancels out upon taking energy differences.

\section{Notes on the simulations}

We work with $M$ rhombically shaped cells stacked on top of each
other in $z$ direction. All of these cells have periodic boundary
conditions in $x$ and $y$ 
directions. Each one of these rhombic units
cells also repeats itself every $M^{th}$ plane due to the periodic
boundary condition in $z$ direction as well. Thus we have periodic boundary
conditions in all directions. There are a total of $N$ pancakes in
each of the $M$ cells. We work with two system sizes to safeguard against any
possibile finite size effects. While working with $N=36$ we
chose $M=25$ in most cases, except for $B=100$ G and $\gamma \leq 150$,
where the number of planes was increased from 25 to 36. For low fields and 
anisotropies the entanglement length along the {\it c} axis becomes
large, and hence in order to observe a sharp transition a larger system
size in the z-direction is needed. 
Similarly, for $64$ pancakes in each plane, we usually
work with a total of 20, 25 or 36 planes depending on the values of
parameter $B$ and $\gamma$. 
In all the cases considered in this paper we always had a total of at least
900 beads ($N=36$, $M=25$) and a maximum of 2304 beads ($N=64$, $M=36$).  
A FL consists of one pancake from each and every plane. Pancakes
belonging to a given FL
were tracked with pointers and linked lists \cite{tildesley}.

Pancakes were moved by either Metropolis algorithm \cite{tildesley} or 
its advance form, multilevel Monte Carlo (MMC) \cite{ceperley}. Which
method to employ in a particular case usually depends upon the
anisotropy and the magnetic field, as described below.

For most anisotropies and magnetic fields, the MMC technique
was used. In MMC we update several pancakes
spread over many planes at once. Thus for the lower anisotropies
$\gamma=125$ and 150, and $B=100$ G,  a total of $15$ to
$20$ beads spread over $5$ planes were used to update the
system. For other anisotropies and fields ($\gamma \geq 250$ and $B
\leq 900$ G) it 
suffices to use just 3 planes in the MMC technique.

For different parameter ranges, one needs to use different methods 
of updating the FL's. For example, one can not use MMC method for
$\gamma=\infty$ since there is no natural choice to generate paths
sampled with a free Gaussian distribution. Even at 
higher values of $\gamma$, such as 500, the MMC method using 5 planes
would be slow and inefficient. Thus for 
$\gamma \geq 375$ and $B \geq 300$, we move only one pancake at
a time using the Metropolis algorithm. 
Flux cutting was implemented by using a different kind of move to
allow for the large wiggling of the FL's (and thus to avoid bias
towards straighter FL configurations). In this move the two ends of 
neighboring lines were switched by considering only the relevant
Josephson interactions without attempting to displace any of the 
pancakes involved. 

Just like the case with high anisotropies discussed above, in the MMC
implementation also we allowed for large wiggling 
of FL's along the $z$-direction by introducing the process of flux 
cutting. Starting
from a few neighboring lines, we cut chunks of FL's spreading over a number
of planes. New paths between the starting and ending positions of the
FL's thus cut were made using a random walk through the space of
permutations \cite{nordborg} and the subsequent use of the
bisection method \cite{ceperley}.

Of course there are ranges of parameters where more than one method of
update can be employed (e.g. Metropolis method, MMC with 3
planes or MMC with 5 planes.). In these cases of overlap, relevant methods
were found to lead to the same result, as they should. 
Further details on the simulation technique are supplied in
Ref. \onlinecite{sandeep}. 

The logarithmic part of the in-plane and out of plane electromagnetic 
interaction (see Eqs.~(\ref{electromagnetic1}) and
(\ref{electromagnetic2})) was 
handled by analytically summing over the interaction as shown in the
Appendix B. The integrals $E_1$ and $E_2$ were evaluated numerically
taking into 
account the periodic boundary conditions in all directions. This is
accomplished by considering images of pancakes in all directions.

\begin{figure}
\includegraphics[width=5in,height=3in]{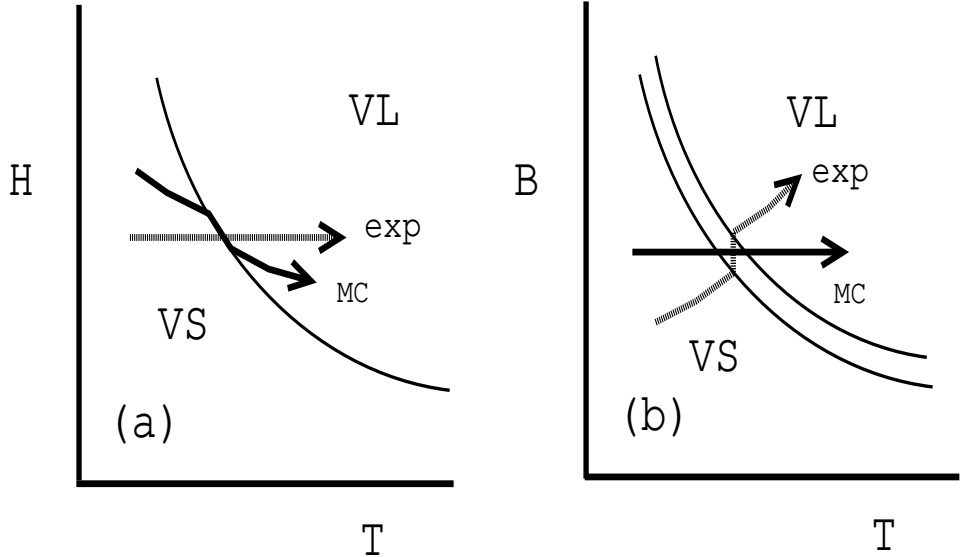}
\caption{Schematic phase diagram showing the vortex-solid (VS) and
  votex-liquid (VL) phases in the $H-T$  and $B-T$ planes. The path the
  system traces as the temperature is raised at fixed $B$ (MC) or
  fixed $H$ (exp) are indicated} 
\label{mcvsexp}
\end{figure}

In the simulations as $T$ is raised it is $B$ that is kept fixed, not
$H$, as done in experiments. This causes the phase transition to be
less sharp, especially at low fields (where the phase boundary is
flatter), as is evident from
Fig. \ref{mcvsexp}. The figure shows both the $H-T$ and $B-T$ phase
diagrams schematically. In the $B-T$ phase diagram there is a region
where both the vortex-solid and vortex-liquid phases coexist. The
paths corresponding to increasing temperature at constant $B$ (MC) and
constant $H$ (experiment) are shown.

The broadening of the phase transition can be avoided by using
`isobaric'' Monte Carlo simulations \cite{nordborg}, but this was not done
in the present work. 

We measured the following physical quantities. For details the reader
is referred to our earlier work \cite{sandeep}.

\subsection{Energy}

An expression for energy can be obtained from

\begin{equation}
E=k T^{2}\frac{\partial }{\partial T}\ln (\Xi (\Lambda ,\beta
,N)).\end{equation} 
Due to the internal temperature dependence of $\lambda$ on $T$, one
gets a very complicated expression for energy (not written down
here). However, a simplified
expression for energy can be obtained under the assumption that $a_0
<\lambda$, 
since in this case it is justified to ignore the internal 
temperature dependence of $\lambda$ on $T$ while taking derivatives with
respect to the temperature \cite{nordborg}. The energy expression
obtained for the case when $a_0 < \lambda$ is given in Ref. 
\onlinecite{sandeep}. This expression, however, was seen to work well
even for the cases where $a_0 \sim \lambda$ and the difference between the
simplified and the exact energy calculations was found to be insignificant for
all cases except for very low values of $B$ such as $40-80$ G (we used
these values for only $\gamma=\infty$). We have
used the simplified expression for the energy in all cases. It would
not affect the melting transition in anyway. The only change will be
that the energies obtained will be off by a few percents for the case
when the magnetic field is very small. 

\subsection{Translational structure factor} 

The translational structure factor $S(\mathbf{Q}_1)$ is defined as, 

\begin{equation}
S(\mathbf{Q}_1)=\frac{1}{MN}\left\langle \sum _{ij,m}e^{\left
(i\mathbf{Q}_1.(\mathbf{R}_{i,m}-\mathbf{R}_{j,m})\right)}\right\rangle,
\label{struc}
\end{equation}
where $\langle ...\rangle $ stands for the MC average, 
and $\mathbf{Q}_1$
stands for a reciprocal lattice vector corresponding to the first
Bragg peak and is given by
\begin{equation}
\mathbf{Q}_1=\frac{2\pi }{a_{0}\sin ^{2}\theta }(\mathbf{e}_{1}-
\mathbf{e}_{2}\cos \theta),
\end{equation}
where $\theta =\pi/3$, $a_{0}$ is the nearest neighbor
distance and $\mathbf{e}_{1,2}$ are the unit vectors along the hexagonal
unit cell such that
\begin{equation}
\mathbf{e}_{1} \cdot \mathbf{e}_{2}=\cos \theta. 
\end{equation}

\subsection{Line entanglement} 

As we allow permutations of FL's, we can define a number $N_{e}/N$
as that fraction of the total number of FL's which belong to loops that
are bigger 
than the size of a ``simple'' loop. A simple loop is defined as a set of $M$
beads connected end to end, $M$ being the total number of planes.
Loops of size $2M$, $3M$... start proliferating at and above the transition
temperature and in the corresponding 2D boson system  this
proliferation is related to the onset of the superfluidity. 

Some of the other important parameters were taken as follows:
$\lambda_0=1700$ {\AA}, $d=15$ {\AA} and $T_c=90$ K.
\section{Results}
\begin{figure}
\begin{center}
\includegraphics{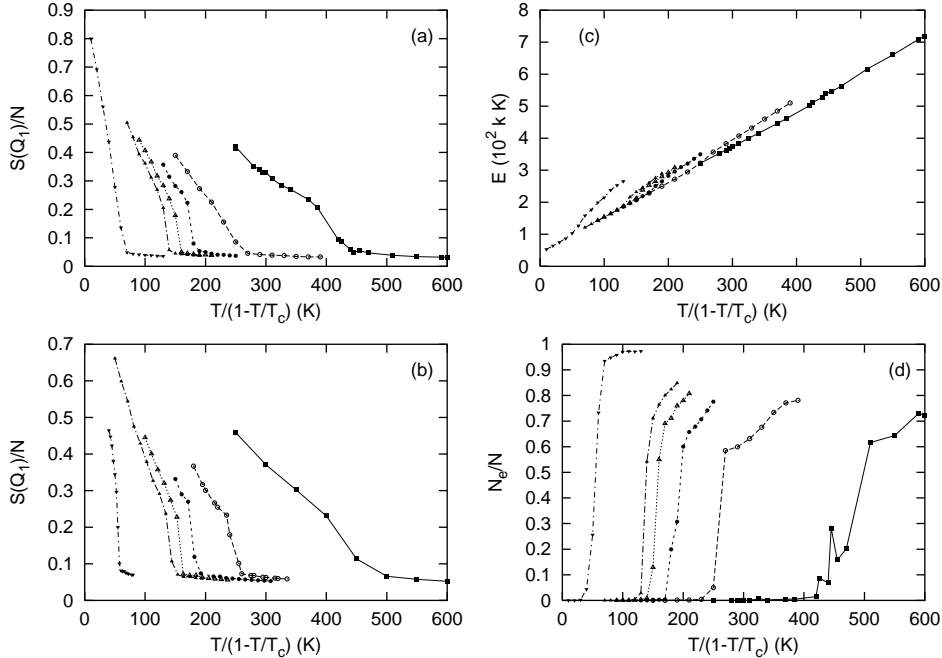}
\end{center}
\caption{$\gamma=125$. For various fields $B=100$ G (filled squares),
$300$ G (open circles), $500$ G (filled circles), $700$ G (open triangles),
$900$ G (filled triangles) and $5000$ G (lower triangles), the following
quantities are shown: 
(a) the translational structure factor at the first Bragg peak for
$N=64$ FL's (b) the translational structure factor at the first Bragg peak for
$N=36$ FL's (c) Energy for $N=64$ FL's (energy is given up to an additive
constant, which is not important.) 
(d) Line entanglement for $N=64$ FL's.}
\label{g125}
\end{figure}

\begin{figure}
\begin{center}
\includegraphics{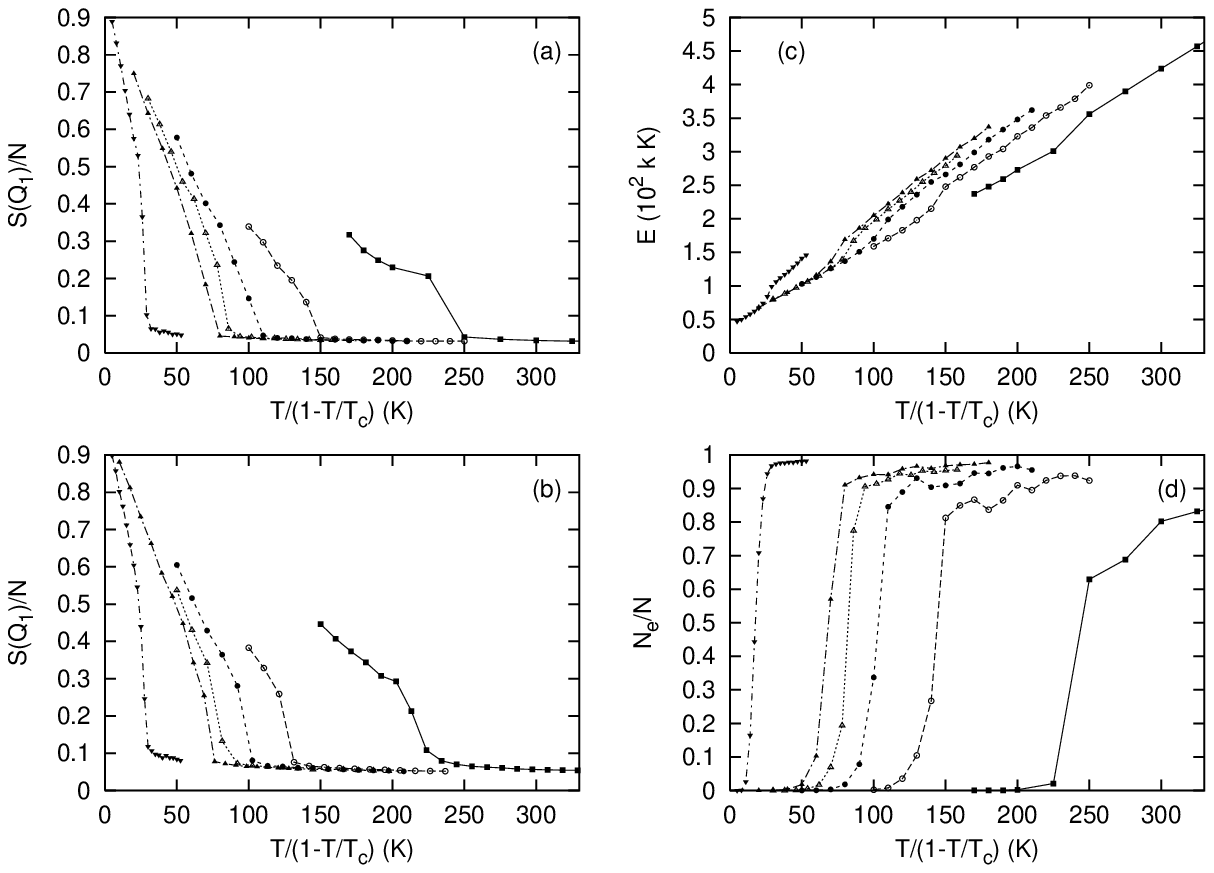}
\end{center}
\caption{$\gamma=250$. The same quantities are shown as in Fig. \ref{g125}
but $\gamma=250$ here.} 
\label{g250}
\end{figure}

\begin{figure}
\begin{center}
\includegraphics{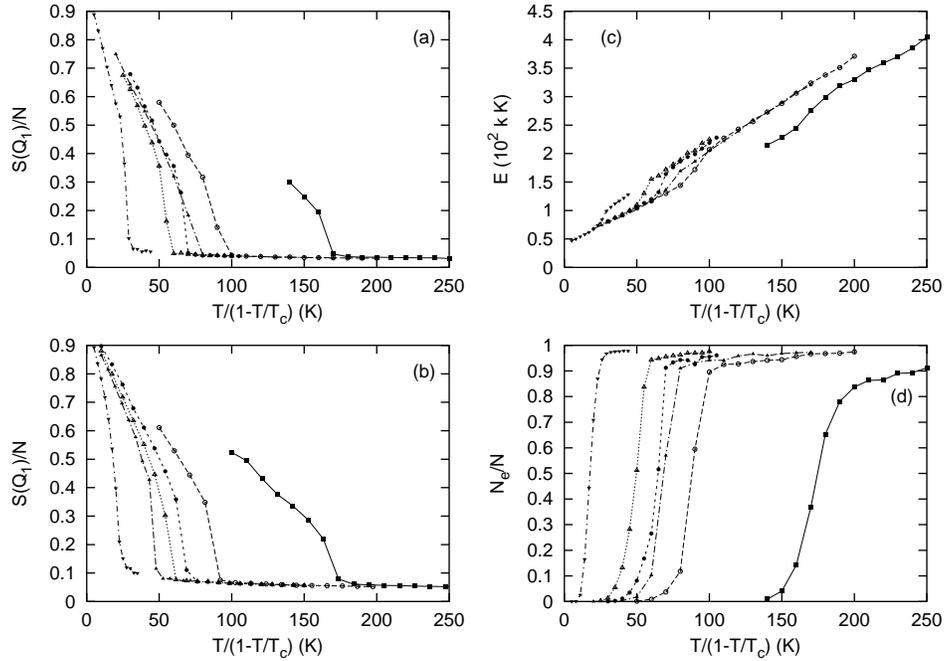}
\end{center}
\caption{$\gamma=375$. The same quantities are shown as in Fig. \ref{g125}
but $\gamma=375$ here.}
\label{g375}
\end{figure}

\begin{figure}
\begin{center}
\includegraphics{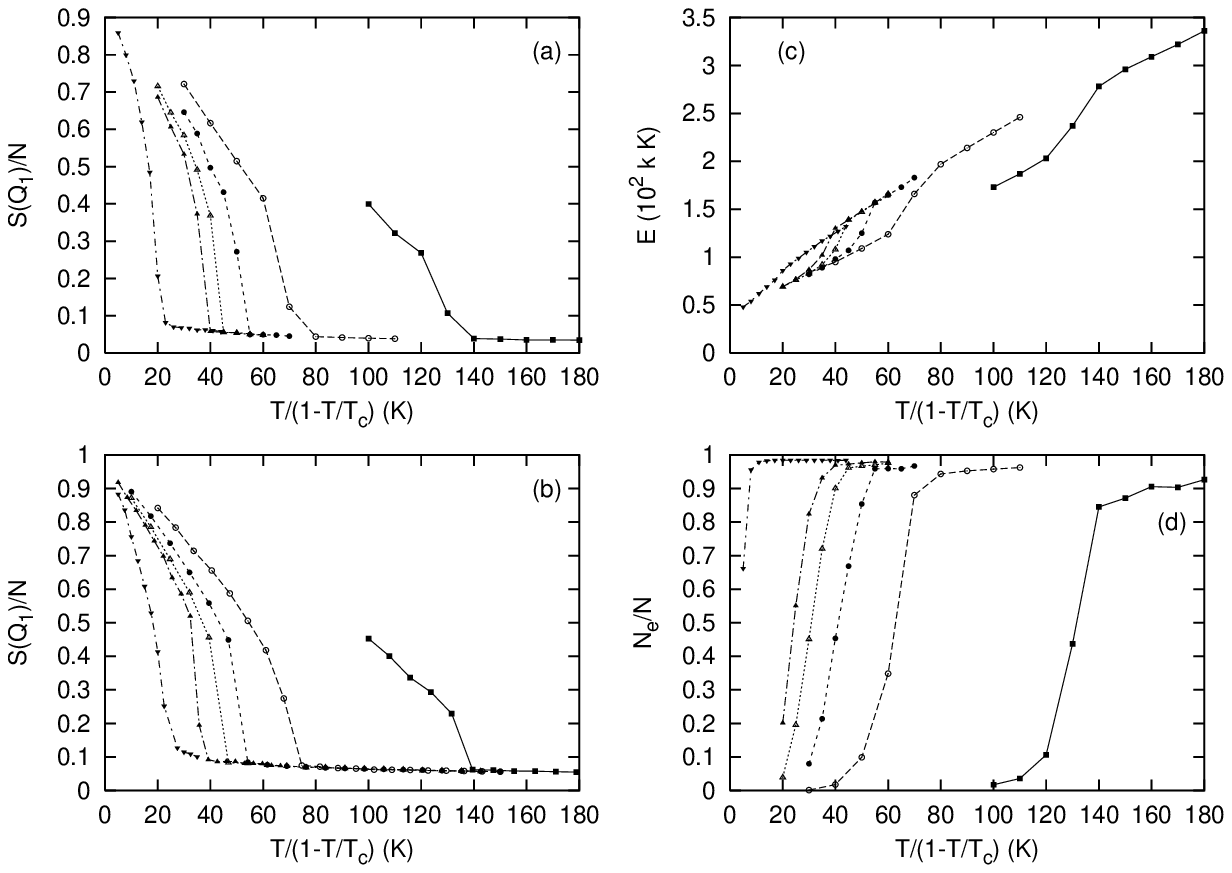}
\end{center}
\caption{$\gamma=500$. The same quantities are shown as in Fig. \ref{g125}
but $\gamma=500$ here.} 
\label{g500}
\end{figure}

\subsection{Josephson and electromagnetic coupling}
In this section we discuss the results when Josephson as well as
electromagnetic couplings are included in the expression for the
free-energy functional. Three different quantities, the translational
structure factor at the first Bragg peak $S(\mathbf{Q}_1)$, the energy 
$E$ and the line entanglement $N_e/N$ were monitored. Simulations were
done for four different anisotropy parameters $\gamma=125, 250, 375$
and $500$.  
In addition we carried out simulations for two different temperature 
dependence of the penetration depth. The results for the first 
temperature dependence of $\lambda$, namely 
$\lambda^2(0)/{\lambda^2(T)}={1-T/T_c}$,
are shown in Figs. (\ref{g125})-(\ref{g500}). The
results for the second dependence of $\lambda$, i.e. 
$\lambda^2(0)/{\lambda^2(T)}={1-(T/T_c)^2}$ 
will be discussed later, in the context of the phase diagram.

To check against possible
finite size effects we worked with two different system sizes, namely
36 and 64 FL's, as
discussed in the previous section. The structure factor
at the first Bragg peak, for the two different sizes is shown in
panels (a) and (c) of Figs. \ref{g125}-\ref{g500}. It is clear that the
transition temperature is unaffected by the choice of the system
size. Similar agreement was seen in plots of energy vs. temperature and the 
graphs of entanglement vs. temperature. These later comparisions
are not displayed.

A first-order transition (FOT) is seen for all anisotropies, which is
in agreement with numerous  experimental \cite{exp1,exp2,exp3,exp4} as
well as numerical \cite{numer1,magro,ryu,koshelev,hu,nordborg,chin,sandeep}
studies on type-II superconductors.
The location of the FOT is inferred from a sharp decay
in $S(\mathbf{Q}_1)$ and a sharp rise in the line
entanglement and a discontinuous jump in energy $E$.  Except for the
case with $\gamma=125$ and $B=100$ G, the transition is sharp and can
be easily located. In all the cases, we see a discontinuous jump in
$E$. The size of the jump can be
determined in the following way. Take two sets of points on $E$
vs. $T_r=T/(1-T/T_c)$ graph, one in the
vortex solid phase and the other in the vortex liquid phase. This can
be done by using $S(\mathbf{Q}_1)$ or $N_e/N$ vs. $T_r$ graphs. We
fit a straight line to the first set of points and
another straight line to the other set of points. The two lines are 
extended up to the transition temperature and the jump in $E$ is read off.

It is clear from the $E$ vs. $T_r$ graphs that for a given anisotropy
one gets lower jumps in $E$ at higher magnetic fields. This effect is
more pronounced for higher anisotropies.

\subsection{Only electromagnetic coupling}
\begin{figure}
\begin{center}
\includegraphics{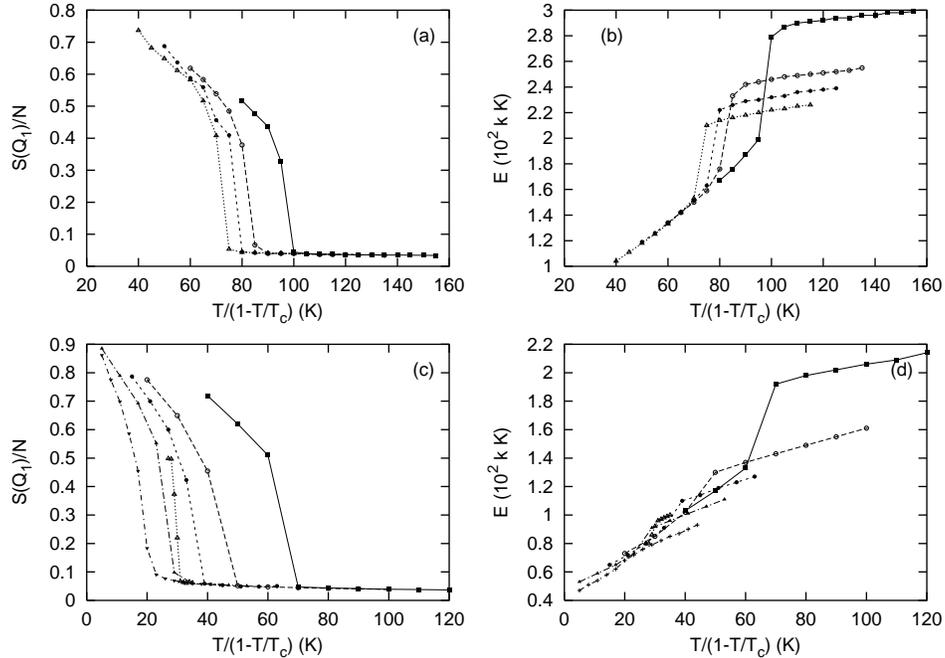}
\end{center}
\caption{$\gamma=\infty$ (no Josephson coupling).
(a) the translational structure factor at the first Bragg peak 
for $B=40$ G (filled squares), $60$ G (open circles), $70$ G (filled
circles), $80$ G (open triangles). (b) Energy for the same fields as in (a)
(c) the translational structure factor at the first Bragg peak for 
$B=100$ G (filled squares), $300$ G (open circles), $500$ G (filled
circles), $700$ G (open triangles), $900$ G (filled triangles)
and $5000$ G (downward triangles). 
(d) Energy for the same fields as in (c).}
\label{ginf}
\end{figure}

Many  simulation  studies completely  neglect  the Josephson  coupling
\cite{olson,fangohr}  and  work   only  with   the  electromagnetic
coupling.  In  this  section  we  discuss the  results  obtained  with
neglecting  the Josephson  interaction ($\gamma=\infty$).  The results
are shown  in Fig.  \ref{ginf}. The  results can be
compared with  a recent simulation    study    of   the    same
system   using    substrate
model \cite{dodgson,fangohr}. The  phase boundary obtained  by Dodgson
{\it et al.} falls almost on top of the phase boundary obtained in this
paper.
Another important feature of this  study is the increasing energy jumps
toward lower  magnetic fields as well as almost diminishing jumps in
$E$ toward very high magnetic fields ($B >900$ G). The
entropy jump calculated using the formula $\Delta s =\Delta  E /T$ shows
an increasing trend towards higher temperature (not shown here) again
in agreement with the references cited above.

Also, for the very high magnetic fields, the transition takes place at
$T_{2D}=16$ K, independent of the magnetic field. 
This result is in agreement with many theoretical
studies \cite{blatter,anisotropy}.  
In this regime of fields each set of pancakes in a plane melts
irrespective of the other planes and
shows a characteristic 2D melting transition at $T_{2D}\approx 16-19$ K.
This kind of 2D melting is expected at high fields as the in-plane 
pancakes get closer
to each other and hence the in-plane interaction keeps
getting stronger compared to out of plane interaction and the
different layers start to melt independent of each other \cite{olson}.

Comparing results with the previous section, it is clear that
the Josephson interaction, present at finite anisotropy, has the
effect of reducing the energy jumps compared to the case when
only the electromagnetic coupling is included.

\subsection{Phase diagram}
\begin{figure}
\begin{center}
\includegraphics{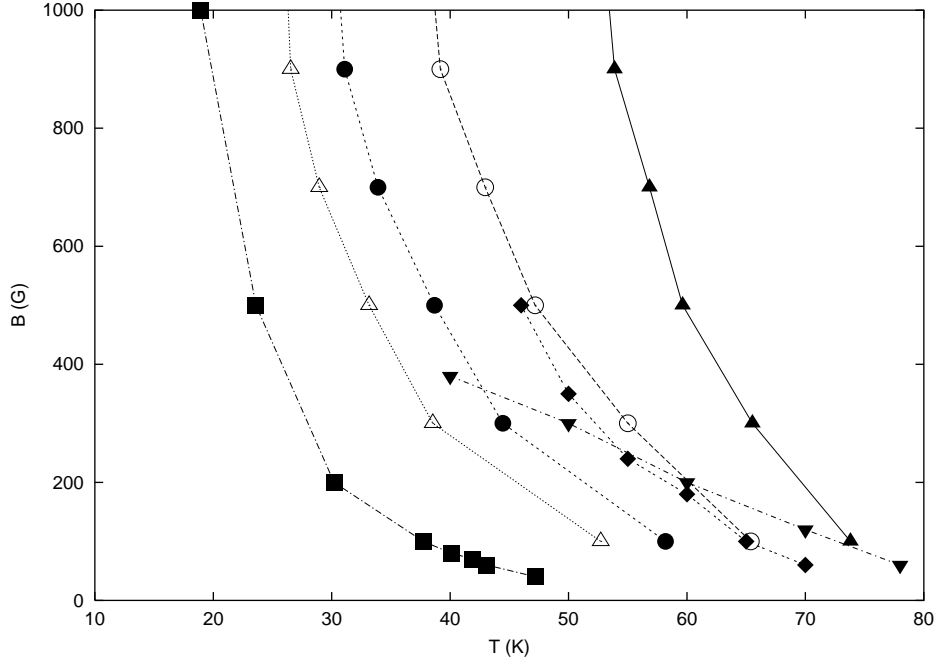}
\end{center}
\caption{Phase diagram obtained by using 
$\lambda(T)=\lambda(0)/\sqrt{1-T/T_c}$:
Melting transitions for various anisotropies
$\gamma=125$ (filled triangles),
$\gamma=250$ (open circles), $375$ (filled circles), $500$ (open triangles),
$\infty$ (filled squares) and experimental(inverted triangles and diamonds) 
are shown.}
\label{phase1}
\end{figure}

Based on the results from the previous two sections, we obtain a phase
diagram for BSCCO which is shown in Fig.~\ref{phase1}. The diagram shows the
melting curves for various anisotropies, including the case when
there is no Josephson coupling. On the same diagram we also show two
experimental lines, one for the melting transition
\cite{exp3,khaykovich} and 
the other for the irreversibility line \cite{zeldov1}. The
experimental irreversibility line falls very close to the simulated 
$\gamma=250$ phase boundary. 
The actual experimental melting curve does not seem to fall on any of
the melting lines obtained from the simulations. The experimental
melting line bends downward toward lower temperatures and
this feature could not be seen here. This may the result of
point defects present even in pristine samples, 
an effect reported in recent simulations \cite{nonomura,olsson}.

\begin{figure}
\begin{center}
\includegraphics{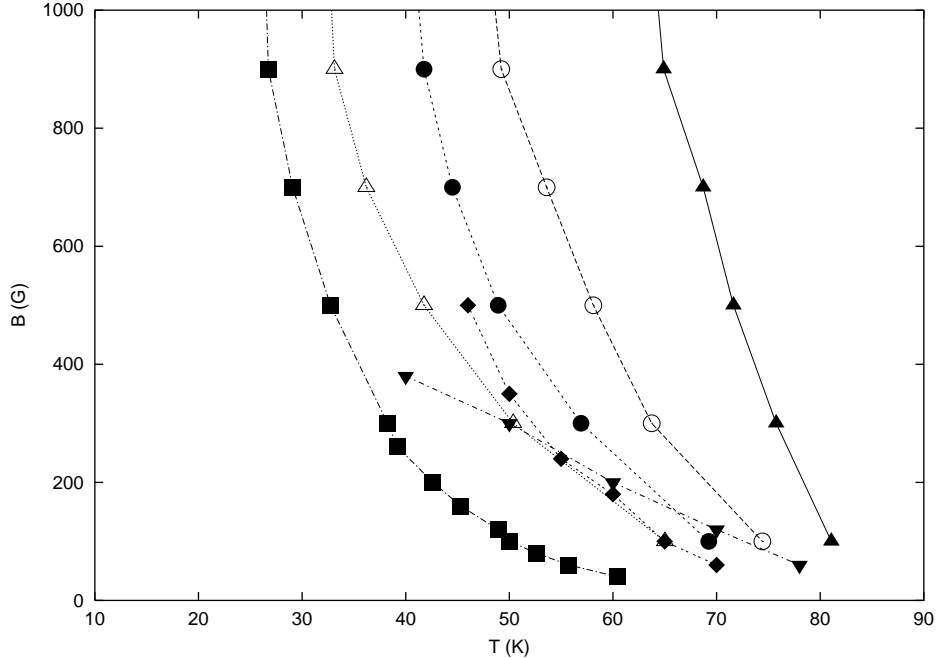}
\end{center}
\caption{ Phase diagram obtained by using 
$\lambda(T)=\lambda(0)/\sqrt{1-T^2/T_c^2}$:
 Melting transitions for various anisotropies
$\gamma=125$ (filled triangles),
$\gamma=250$ (hollow circles), $375$ (filled circles), $500$ (hollow
triangles), 
$\infty$ (filled squares) and experimental melting transition
(inverted triangles) and the experimental irreversiblity line (diamonds) 
are shown.}
\label{phase2}
\end{figure}

In Fig. \ref{phase2} we show the phase diagram obtained by 
assuming a different dependence of $\lambda$ on temperature, namely 
$\lambda(T)=\lambda(0)/\sqrt{1-T^2/T_c^2}$. For this choice the irreversibility
line falls somewhat below the $\gamma=500$ simulated phase boundary,
leading to an estimate of $\gamma \approx 450$. As discussed above the
temperature dependence should be somewhere in between these
dependences so an estimate of $\gamma=300-400$ for the experimental
sample is probably reasonable.

\begin{figure}
\begin{center}
\includegraphics{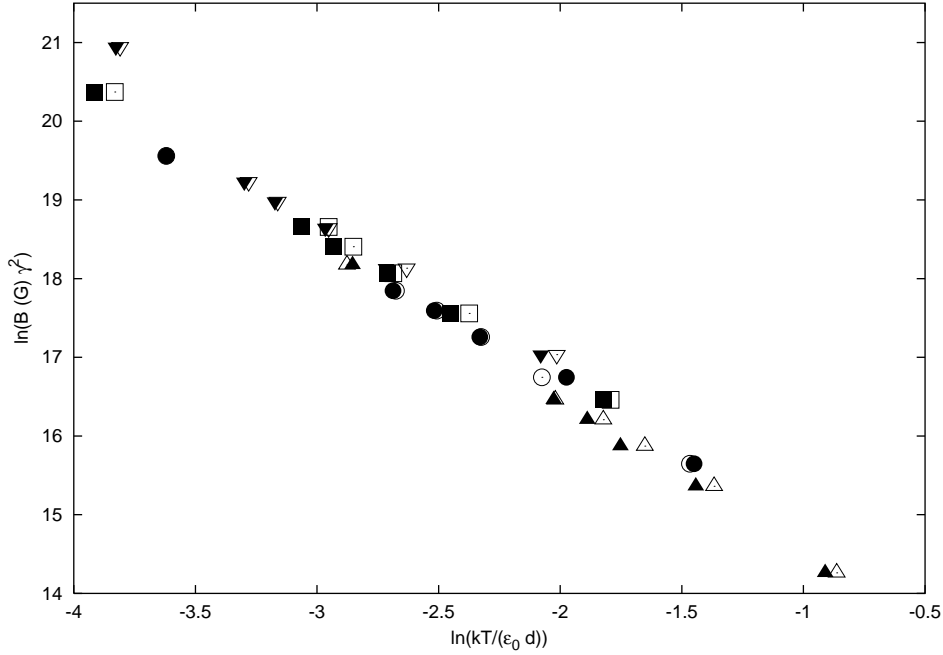}
\label{colapselog1}
\end{center}
\caption{Data for finite anisotropies $\gamma=125$ (
  triangles),$250$ (circles),$375$ (squares) and $500$ (inverted triangles) 
from Fig.~\ref{phase1} and Fig.~\ref{phase2} collapse
 onto a single curve. Filled symbols are for the $
\lambda^2(T)=\lambda^2(0)/(1-T/T_c)$ dependence while hollow symbols
are for $\lambda^2(T)=\lambda^2(0)/(1-T^2/T_c^2)$ dependence. 
Some spread in the data is attributed to the fact
that this scaling is not valid at high values of anisotropies where
Josephson coupling becomes comparable to the electromagnetic coupling.}
\label{collow}
\end{figure}

We have found that the data for the phase boundaries can be collapsed
on a single curve for the case of finite anisotropy even when using
different temperature dependences of $\lambda$. When we plot the
variable $B\gamma^2$ at the melting transition versus the variable 
$kT/(\varepsilon_0(T) d)$ all 8 curves fall on top of each
other. Koshelev  \cite{koshelev} argues that when Josephson coupling
dominates over electromagnetic coupling the phase boudary is determined
by a single dimensionless function of the dimensionless parameters
$kT/\varepsilon_0 d$ where
$\varepsilon_0(T)=\phi_0^2/(4\pi\lambda(T))^2$ and $r_g^2/a_0^2\propto
{B\gamma^2}$. In Fig.(\ref{collow} we plot 
$\ln (B\gamma^2)$ versus $\ln (kT/(\varepsilon_0 d))$ and the data
collapses to a straight line with slope $(-2)$. This suggests that the
transition is given by a single relation 
\begin{eqnarray}
 \Lambda \equiv\left(\frac{kT}{\sqrt{2}\varepsilon_0 d}\right)
\left(\frac{\gamma d}{a_0}\right)=\Lambda_c
\label{Lambda} 
\end{eqnarray}

as is known to be the case for YBCO \cite{nordborg}. From a simple
cage model (see e.g. Ref. \onlinecite{crabtree}) it follows that the
$\Lambda_c=c_L^2$, where $c_L$ is the Lindemann coefficient. This
means that the transition occurs when the mean square devaiation of
the FL from a straight line exceeds $c_L a_0$, where $a_0$ is the
lattice constant. Since we find that $\Lambda_c\approx 0.1$, it appears 
that $c_L\approx 0.3$ (It was estimated to be 0.25 for YBCO \cite{nordborg}).
Notice that actually the expression for $\Lambda$ should be $kT/(a_0
\sqrt{2 \varepsilon_l \varepsilon_0})$ and 
we have used $\varepsilon_l=\varepsilon_0/\gamma^2$. Had we included
the factor $\ln(\lambda/d)$ in $\varepsilon_l$ we would have obtained
$\Lambda_c=0.1/\sqrt{5}\approx 0.045$ and $c_L \approx 0.2$. Thus
$c_L$ turns out to be comparable to that found for YBCO. 
This phase diagram shows the importance of keeping the Josephson
coupling. Even if the anisotropy parameter is as large as
$\gamma=500$, it is still not appropriate to neglect the
Josephson coupling entirely. 

\begin{figure}
\begin{center}
\includegraphics{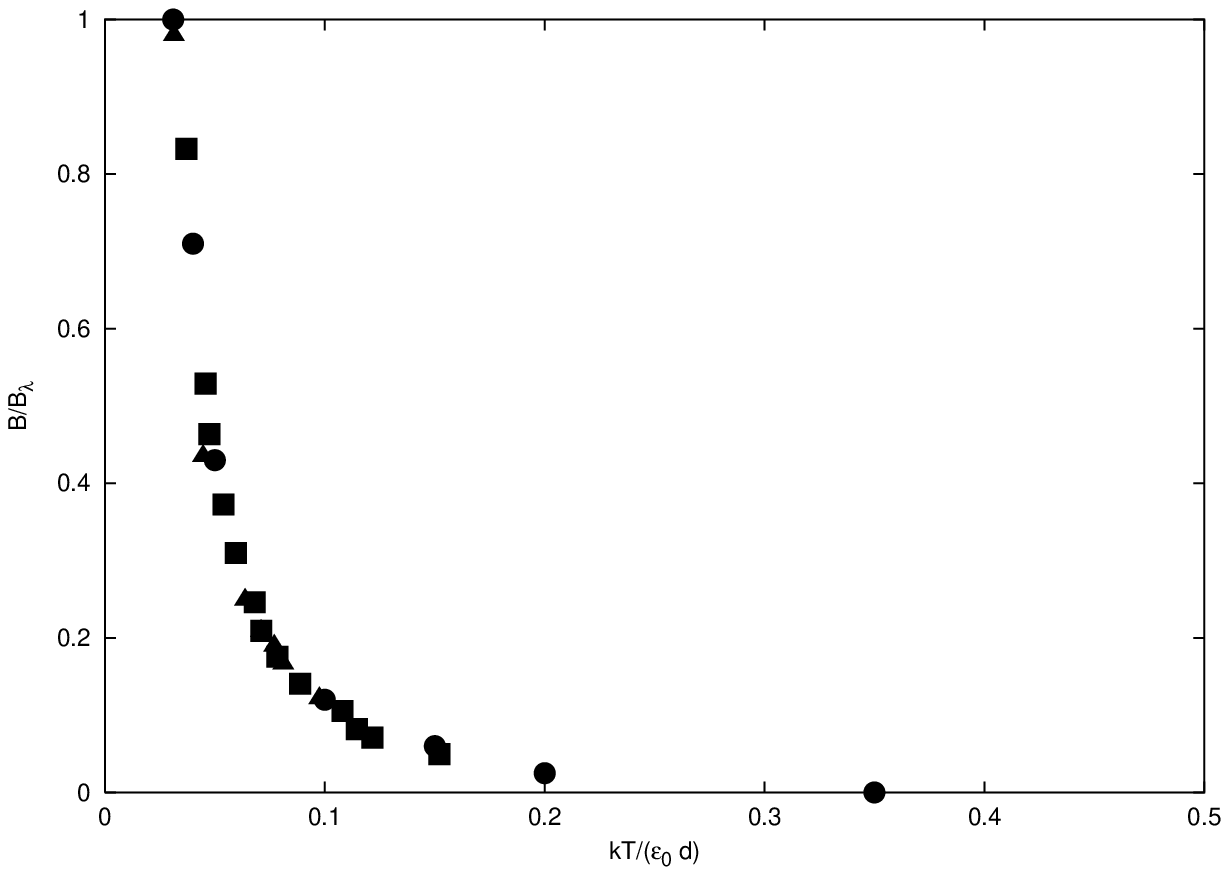}
\label{colapse2}
\end{center}
\caption{Data for $\gamma=\infty$ from Fig.~\ref{phase1} (triangles) and 
Fig.~\ref{phase2} (squares) collapse
onto a single curve. Also shown is the melting line obtained by Dodgson
{\it et al.} \cite{dodgson} (circles).}
\label{colinf}
\end{figure}

For $\gamma=\infty$ there is another scaling which collapses the data 
\cite{dodgson,fangohr}
for the two kinds of temperature dependence we considered, namely
plotting $B/B_\lambda$ at the melting transition vs. $kT/(\varepsilon_0
d)$, where $B_\lambda=\phi_0/\lambda^2(T)$.
This is an approximate scaling of the electromagnetic interaction
valid when the dependence on the small parameter $d/\lambda$ can be
neglected. In Fig. (\ref{colinf}) we show the collapsed data for two different
temperature dependences and also compare with the result of Dodgson
{\it et al} \cite{dodgson,fangohr}.

\subsection{Comparison: local and nonlocal interactions}

\begin{figure}
\begin{center}
\includegraphics{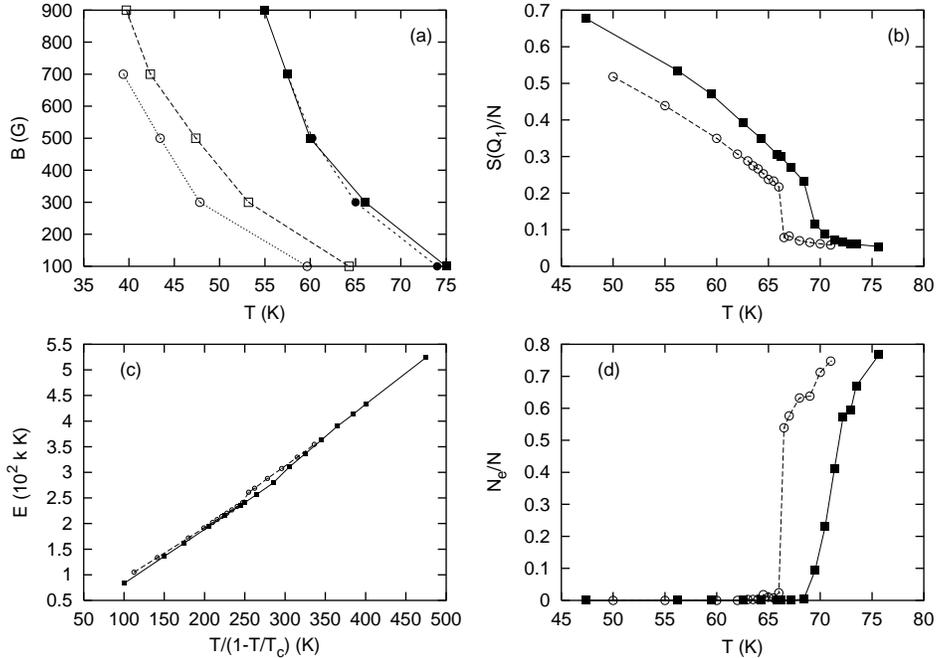}
\end{center}
\caption{ Comparison between two models:
(a) Phase boundaries for two different models, this paper (squares),
Ref. \onlinecite{sandeep} (circles) for anisotropy $\gamma=125$ (solid
squares and circles) and $\gamma=250$ (blank squares and circles). (b) 
$S(\mathbf{Q}_1)$
vs. temperature for $\gamma=150$. this paper(solid squares),
Ref. \onlinecite{sandeep}(open circles) (c) and (d) show $E$ and $N_e/N$
with the same symbols and $\gamma$ value as in part (b)}
\label{comparison}
\end{figure}

It is easy to show that the expression for the interaction of a single
pancake with an infinite set of pancakes stacked on
top of each other reduces to the modified Bessel function of second
kind $K_0(R/\lambda)$ where $R$ is the distance of the pancake
from the stack of the pancakes (see Appendix A). At lower
temperatures, magnetic field 
and anisotropy, the FL's are straight so it has been considered a good
approximation to replace the nonlocal interaction between the pancakes
with a local one of type mentioned above. For example we have used
this model in a recent simulation \cite{sandeep}.
In this section we compare the results obtained by replacing the
nonlocal electromagnetic interaction with a local interaction of the
$K_0$ type. The results are shown
in Fig. \ref{comparison}, assuming the temperature dependence
$\lambda^2(0)/\lambda^2(T)=1-t$. In Fig. \ref{comparison}(a) we show the
phase lines for  
$\gamma=125$ and $250$. The phase lines for $\gamma=125$ from the two
models fall 
almost on top of each other. Even for $\gamma=250$ the deviation
is still small. This is reasonable, as one expects that the FL
deviates less from a
straight line configuration when $\gamma$ is small.
In Fig. \ref{comparison}(b-d) we compare the results obtained here with a
previous paper \cite{sandeep} for $\gamma=125$ and $B=125$ G. There is a small
shift in transition temperature of about $3$ K. It is clear that if
one shifts the $S(\mathbf{Q}_1)$ line obtained with
$K_0(R/\lambda)$ interaction by around $3$ K to the right, then the
two curves will fall almost on top of each other.

Fig. \ref{comparison}(b) compares the energy jumps for the same two case as
in Fig. \ref{comparison}(b). Except for slight shift in the transition
temperature, the jumps in energy are
identical. Fig. \ref{comparison}(c) shows the 
line entanglement and as expected we again see a slight shift in the
position where line entanglement takes place, consistent with
Fig. \ref{comparison}(b) and \ref{comparison}(c).

This justifies our claim that for the smaller anisotropies ($\gamma
\approx 125$) 
it is a good approximation to replace the non local
electromagnetic interaction with a local one of the type
$K_0(R/\lambda)$.

\section{Conclusions}

In this work we introduced a model which incorporates the
electromagnetic interaction and Josephson interaction among 
pancake vortices in the highly anisotropic BSCCO in a more 
systematic way than previously done. Instead of approximating the effect of the
electromagnetic interaction by an in-plane repulsive interaction
between pancakes that is given by a Bessel function of the second
kind we treat this interaction exactly and in addition we include the
Josephson coupling.
Treating the electromagnetic interaction as an effective Bessel
function interaction is valid only if FL's do not deviate too
much from straight lines. 
These distortions can be quite large close to the melting transition
when the flux cutting mechanism starts to proliferate. Thus in the
present model interaction among pancakes is taken in a more realistic
manner. There are still some approximations involved in this model but
we believe that the result are more acurate than previousely obtained.

The phase boundaries for various anisotropies were obtained. It was shown
that for finite anisotropy, when Josephson coupling plays a role, it
tends to smoothen out energy jumps compared to the case when only the 
electromagnetic coupling is kept. Thus energy jumps tend to increase
with increasing anisotropy. They are also found to decrease with
increasing magnetic field along the phase boundary.
The phase diagram clearly shows a shift in the transition temperature 
as a function of the anisotropy.

For finite anisotropy up to a value of 500 we showed that the data can
be collapsed to a straight line in the ln-ln plot of $B\gamma^2$
versus $kT/\varepsilon_0 d$ meaning that the transition occurs when
a single variable combination of temperature, magnetic field and
anisotropy becomes critical. From here we could deduce the value of
the Lindemann parameter to be about 0.3. For infinite anisotropy we
obtained scaling when plotting $B\lambda^2/\phi_0$ vs. $kT/\varepsilon_0 d$.

While keeping just the electromagnetic coupling we observe a typical 2D
melting behavior towards high magnetic fields where the transition
temperature $T_{2D} \sim 16$ K becomes independent of the magnetic
field. Our results for  
$\gamma=\infty$ were compared with a recent simulation
study using the substrate model \cite{dodgson,fangohr}. The
agreement between the two results is excellent.

Finally we have compared the two
cases where the electromagnetic interaction is treated approximately
(Ref. \onlinecite{sandeep}) and exactly (this paper) for a value of
$\gamma=125$. It is shown that in this case
transition temperatures are very close and keeping just the in-plane
interaction in the form of a 
modified Bessel function of the second kind is a good approximation, 
justifying the claim that we have made in a previous
paper \cite{sandeep}. In that paper our main goal was to extract the
effect of columnar pins on the melting transition. For larger 
anisotropies the deviations of the two treatments become more pronounced.

Comparing with experimental results the simulations seem to agree
better with the position of the so-called irreversibility line than
with the position of the true melting line. This may be due to the
effect of point defects in the experimental samples. Even pristine
sample contains point defects that tend to 
shift the phase boundary towards lower temperatures and flatten it out
somewhat at the low temperature side. Point defects can also have an
effect on the energy and entropy jumps across the phase boundary.
If one would like to extract the value of the anisotropy of the
experimental sample from the coparison with the present simulations,
we would estimate it to be in the 250-450 range.

\section{Acknowledgments}
This work is supported by the US Department of Energy (DOE), Grant 
No. DE-FG02-98ER45686. YYG also thanks the Kavli Institute for
Theoretical Physics (KITP) in Santa Barbara for its hospitality, where
part of this paper has been finalized.

\newpage
\appendix

\section{Straight Flux-line Approximation}
In this section we show that the interaction of a pancake with a stack of
pancakes  located at a distance $R$ away from the given pancake and
lined up along the $z$ direction reduces to the modified Bessel function
of the second kind, $K_{0}(R/\lambda)$.

Consider the interaction of a pancake situated at z=0 with a stack of
pancakes. We 
begin with Eq. (\ref{electromagnetic1}) and (\ref{electromagnetic2}). The
total interaction of the pancake with the FL will be given by

\begin{eqnarray}
{\bf U}_{total}\left(\frac{R}{\lambda}\right)=2\,d \,\varepsilon_{0}
\ln{\frac{C}{R}}+Y_1\left(\frac{R}{\lambda}\right)
+Y_2\left(\frac{R}{\lambda}\right) 
\label{append2.2},
\end{eqnarray}
where $Y_1(R/\lambda)$ and $Y_2(R/\lambda)$ are given by

\begin{eqnarray}
Y_1\left(\frac{R}{\lambda}\right)=-\frac{d^{2}\varepsilon_{0}}{\lambda}
\left(\sum_{m=-\infty}^{+\infty}\exp(-|m \,d|/\lambda)\right)\ln\frac{C}{R} , 
\label{append2.3}
\end{eqnarray} 
and,

\begin{eqnarray}
Y_2\left(\frac{R}{\lambda}\right)=\frac{d^{2}\varepsilon_{0}}{\lambda}
\left(\sum_{m=-\infty}^{+\infty}\int_{R}^{\infty}
dR^{\prime}\,\frac{\exp(\,-\sqrt{|m \,d|^2+
R^{\prime 2}}/\lambda\,)}{R^{\prime}}\right).
\label{append2.4}
\end{eqnarray} 

We have a geometric series in Eq. (\ref{append2.3}) which can easily be
summed over to give

\begin{eqnarray}
Y_1\left(\frac{R}{\lambda}\right)=-\frac{d^{2}\varepsilon_{0}}
{\lambda}\left(\frac{1+\exp(-d/\lambda)}
{1-\exp(-d/\lambda)}\right)\ln\frac{C}{R} . 
\label{append2.5} 
\end{eqnarray}
As $\lambda/d \sim 120$ so we can approximate Eq.(\ref{append2.5}) with 

\begin{eqnarray}
Y_1\left(\frac{R}{\lambda}\right)=-2 \,d \,\varepsilon_{0}\,\ln\frac{C}{R} , 
\label{append2.6}. 
\end{eqnarray}
Replacing the summation over $m$ to an integration over $z$ and changing the
order of integration we can express $Y_2\left(R/\lambda\right)$ as

\begin{eqnarray}
Y_2\left(\frac{R}{\lambda}\right)=\frac{d \,\varepsilon_{0}}{\lambda}
\left(\int_{R}^{\infty}dR^{\prime}\,\int_{-\infty}^{\infty}
  dz\,\frac{\exp(\,-\sqrt{z^2+ 
R^{\prime 2}}/\lambda\,)}{R^{\prime}}\right).
\label{append2.7}
\end{eqnarray}
With a simple change of variable over z, $ z=R^{\prime}\,t $ and
then again exchanging the order of integration we get

\begin{eqnarray}
Y_2\left(\frac{R}{\lambda}\right)=\frac{d \,\varepsilon_{0}}{\lambda}
\left(\int_{-\infty}^{+\infty} dt\,\int_{R}^{\infty}
dR^{\prime}\,\exp(\,-R^{\prime}\sqrt{1+
t^2}/\lambda\,)\right).
\label{append2.8}
\end{eqnarray}
 
Finally integrating over $R^{\prime}$ one obtains
\begin{eqnarray}
Y_2\left(\frac{R}{\lambda}\right)&=& d \,\varepsilon_{0}
\left(\int_{-\infty}^{\infty} dt\,\frac{\exp(\,-R \sqrt{1+
t^2}/\lambda\,)}{\sqrt{1+t^2}}\right) \\
          &=&2 \,d\, \varepsilon_{0} \,K_0\left(\frac{R}{\lambda}\right).
\label{append2.9}
\end{eqnarray} 
Combining Eq. (\ref{append2.2}), (\ref{append2.6}) and
(\ref{append2.9}) we get
\begin{equation}
{\bf U}_{total}\left(\frac{R}{\lambda}\right)=2\,d
\,\varepsilon_{0}\,K_0\left(\frac{R}{\lambda}\right). 
\label{append2.10}
\end{equation}
\section{Energy sum over the images}

We again consider a rhombically shaped region with side $L$ and angle
$\theta ,$ 
unit vectors are $\mathbf{e}_{1}$ and $\mathbf{e}_{2}$  with 
$\mathbf{e}_{1}\cdot\mathbf{e}_{2}=\cos \theta$, as was done in the appendix
of a previous paper. \cite{sandeep} The Green's function
$G_{0}$ which describe the 2D coulomb interaction between one vortex
and another including all its images, as is implied by the periodic
boundary conditions is given by the solution to London's equation 
(see e.g. Ref.
\onlinecite{tinkham})

\begin{equation}
(1-\lambda ^{2}\nabla ^{2})G_{0}(\mathbf{R},\lambda )=2 \pi \lambda ^{2}\delta 
(\mathbf{R}),
\end{equation}
 with the parameter $\lambda $ setting the scale for the range of
the interaction. The solution of the above equation was derived in the 
appendix of Ref.\onlinecite{sandeep}. The result was
\begin{equation}
G_{0}(\mathbf{R},\lambda )=\frac{\sin {\theta }}{2}\sum _{n=
-\infty }^{+\infty }\frac{\cos (t_{2}n-2\pi \beta _{n})\sinh 
(\gamma _{n}t_{1})+\cos (t_{2}n)\sinh (\gamma _{n}(2\pi -t_{1}))}
{\gamma _{n}(\cosh (2\pi \gamma _{n})-\cos (2\pi \beta _{n}))},
\end{equation}
 where
\begin{equation}
t_{1}=\frac{2{\pi }R_{1}}{L},\, \, t_{2}=\frac{2{\pi }}{L}(R_{1}
\cos \theta +R_{2}),\, \, \beta _{n}=n\cos \theta ,\, \, 
\gamma _{n}=\sin \theta \sqrt{n^{2}+L^2/(2\pi \lambda)^2}.
\end{equation}
To get a formula for Logarithmic interaction all we have to do is to take 
the limit $\lambda$ tends to infinity. In this limit there is only one
term which diverges in the series given above namely that corresponding to 
$n=0$. Separating out the term corresponding to $n=0$ we get

\begin{equation}
G_{0}(\mathbf{R},\lambda )=term_{0}+
\sin \theta\sum _{n=
1 }^{+\infty }\frac{\cos (t_{2}n-2\pi \beta _{n})\sinh 
(\gamma _{n}t_{1})+\cos (t_{2}n)\sinh (\gamma _{n}(2\pi -t_{1}))}
{\gamma _{n}(\cosh (2\pi \gamma _{n})-\cos (2\pi \beta _{n}))},
\label{B4}
\end{equation}
where,

\begin{equation}
term_{0}=\frac{\sin\theta}{2}
\frac{\sinh 
(\gamma _{0}t_{1})+\sinh (\gamma _{0}(2\pi -t_{1}))}
{\gamma _{0}(\cosh (2\pi \gamma _{0})-1)}.
\end{equation}

In the equation above  $\gamma_{0}=L/(2 \pi \lambda)$ tends to zero. 
Taking the limit $ \gamma_{0} \rightarrow 0$ gives

\begin{eqnarray}
term_{0}= \frac{\sin\theta}{2} \left[
\frac{1}{\pi \gamma_0^2}+\frac{1}{3} \pi\left(1-6\left(\frac{R_1}{L}\right)+
6\left(\frac{R_1}{L}\right)^2\right)\right].
\end{eqnarray}

Now the first term in the equation above is a constant (albeit
infinite) independent of $R_1$ and $R_2$ and after dropping it
we are left with the following expression:

\begin{eqnarray}
term_{0}
= \frac{\sin\theta}{2} \left[
\frac{1}{3} \pi\left(1-6\left(\frac{R_1}{L}\right)+
6\left(\frac{R_1}{L}\right)^2\right)\right].
\label{B7}
\end{eqnarray}
Eq. (\ref{B4}) with with $\gamma_n=n \sin\theta$ together with Eq. (\ref{B7})
constitute the required summation over a logarithmic potential under
2D periodic boundary conditions.

\end{document}